\documentclass[letterpaper, 10 pt, conference]{ieeeconf}  

\IEEEoverridecommandlockouts                              

\usepackage{xcolor}
\usepackage{caption}
\usepackage{amssymb}
\usepackage{graphicx}
\usepackage{caption}
\usepackage{algorithm}
\usepackage{algorithmicx}
\usepackage{algpseudocode} 
\usepackage{amsmath}
\usepackage{booktabs}
\usepackage{subcaption}
\usepackage{comment}
\usepackage{wrapfig}
\usepackage{float}
\usepackage{setspace}

\usepackage[most]{tcolorbox}
\usepackage[top=57pt,bottom=43pt,left=48pt,right=48pt]{geometry}

\newcommand{\cmark}{$\bigcirc$} 
\newcommand{\xmark}{$\times$}   

\title{\LARGE \bf
Reachability-based Temporal Logic Verification for\\
Reliable LLM-guided Human-Autonomy Teaming}

\author{{Joonwon~Choi, Kartik~Anand~Pant, Karthik~Nune, and Inseok~Hwang}
\thanks{J. Choi, K. Pant, and I. Hwang were with the School of Aeronautics and Astronautics, Purdue University, Indiana, IN, 47907 USA e-mail: (choi774, kpant, ihwang)@purdue.edu.}
\thanks{K. Nune was with the Interdisciplinary Engineering, Purdue University, Indiana, IN, 47907 USA e-mail: knune@purdue.edu}
}

\begin{document}

\maketitle
\thispagestyle{empty}
\pagestyle{empty}

\begin{abstract}

We propose a reachability-based framework for reliable LLM-guided human-autonomy teaming (HAT) using signal temporal logic (STL). In the proposed framework, LLM is leveraged as a translator that transfers natural language commands given by a human operator into corresponding STL specifications or vice versa. An STL feasibility filter (SFF) is proposed to check the feasibility of the generated STL. The SFF first decomposes the complex and nested LLM translation into a set of simpler subformulas for parallelization and informative feedback generation. The reachability analysis method is then applied to verify if each subformula is feasible for a target dynamical system: if feasible, perform mission planning, otherwise, reject it. The proposed SFF can identify infeasible subformulas, more than simply providing the boolean verification results for the whole STL, thereby facilitating the feedback generation of LLM to request modification of the command to the human. Consequently, the proposed framework can allow more reliable HAT by enabling safe and informative communication between the human operator and the autonomous agent. Our experiments demonstrate that the proposed framework can successfully filter out infeasible subformulas and generate informative feedback based on such information. 
\end{abstract}


\section{Introduction}

Human-autonomy teaming (HAT) is teamwork between humans and autonomous agents, where the autonomous agents are entities with a partial or high degree of self-governance with respect to decision-making, adaptation, and communication \cite{o2022human}. 
If well-designed, a HAT framework can outperform human-only and machine-only cases by complementing each other's limitations~\cite{cummings2014man, lyons2021human}.
To this end, informative and transparent communication between the human and the autonomous agent is crucial~\cite{lyons2021human, gruber2024impact}. Several natural language processing (NLP)-based frameworks have been proposed to facilitate the understanding of autonomous agents on human command.
Particularly, the recent advances in large-language models (LLMs) have found their utility in HAT settings, specifically in prompting and assisting humans based on LLMs' strength in NLP~\cite{liu2024effect, tanneberg2024help, jin2022integrating, wang2023conformal}. 

However, using LLMs for applications beyond language modeling and translation to actual robot operations often raises concerns about safety and reliability~\cite{wu2024safety}, particularly for safety-critical scenarios like HAT.
Despite considerable successes in utilizing LLMs for HAT applications, relatively few works have formally guaranteed the safety or feasibility of the LLM's behavior~\cite{hafez2025safe}. 
Some early efforts have been in this direction to incorporate the safety property while using LLMs for planning and control~\cite{wang2023conformal, hafez2025safe, wu2024selp, wang2024ensuring}. 
However, these methods perform a post-verification (verification during or after planning/execution) as opposed to a pre-verification (verification before planning/execution). It is desirable to have frameworks that allow the robots to perform verification before a mission planning, as well as a mechanism that ensures a semantic feedback is provided back to the human to enhance their awareness.

\begin{figure*}[t]
    \centering
    \includegraphics[width=\textwidth, trim=7 0 0 0, clip]{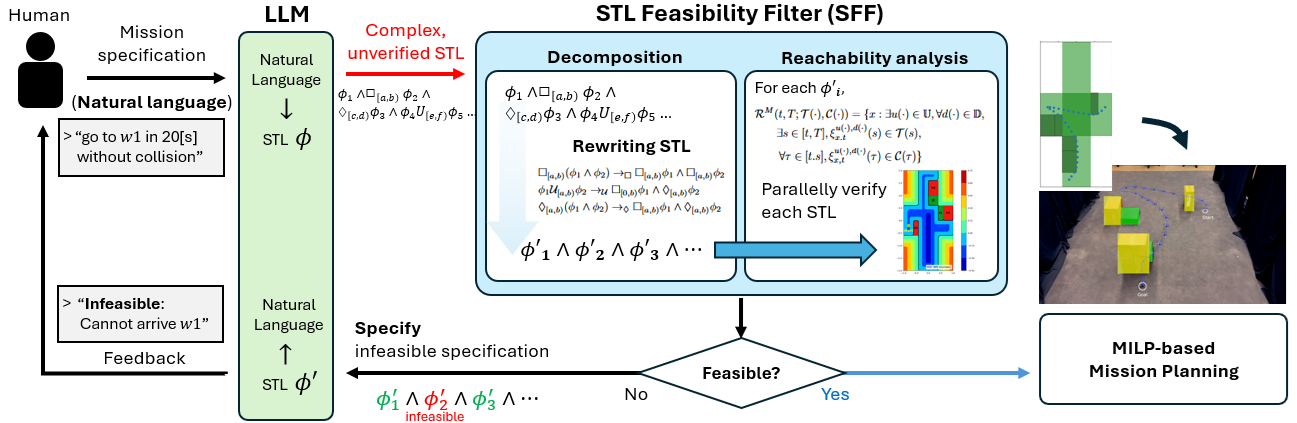} 
    \caption{Architecture of the proposed framework}
    \label{intro_fig}
\end{figure*}

Motivated by the problems mentioned above, we propose a reliable LLM-guided HAT framework that promotes informative communication between the human and autonomous agents. The overall architecture of the proposed framework can be found in Fig.~\ref{intro_fig}. In the proposed framework, we focus on the LLM's strength in NLP~\cite{brown2020language} along with reasoning~\cite{matarazzo2025survey}; and utilize LLM as a translator to translate the human's natural language commands into a corresponding signal temporal logic (STL) specification. The generated STL then passes through the proposed STL feasibility filter (SFF) which verifies its feasibility by a target autonomous system, i.e., if the agent can perform the mission while satisfying the specifications. 

The SFF first decomposes the STL into several subformulas based on~\cite{leahy2023rewrite}, to achieve faster computation via parallelization and identification of infeasible subformulas. 
Then, we leverage the reachability analysis to check each subformula. The reachability analysis is a method that considers all the reachable states of a target system, enabling rigorous examination of future trajectories of an autonomous system~\cite{choi2024data}. 
Particularly, in~\cite{chen2018signal}, the correspondence between the STL and reachability analysis was investigated, showing the satisfaction of STL can be investigated through the reachability analysis formulation. Accordingly, one can verify the feasibility of the decomposed STL based on the reachability. 

If the mission is verified to be feasible, the proposed algorithm starts mission trajectory planning. Otherwise, the identified infeasible subformulas are delivered to LLM to generate informative feedback that explains why the mission is infeasible. Therefore, the proposed framework can formally verify the mission while facilitating the informative communication within HAT, rigorously checking both the agent's physical capability to perform the mission and LLM's uncertainties. 
Consequently, the proposed SFF has the benefit of generating informative feedback by leveraging such infeasible STL information as opposed to providing a boolean result (feasible or not).

We evaluate the proposed framework using several available LLMs and use comparative scenarios to validate its performance rigorously. We observe that the proposed SFF can successfully identify infeasible subformulas from LLM's output, as opposed to simply providing a Boolean output for feasibility and safety. Furthermore, the decomposition scheme also shows a benefit in reducing the computational burden in MILP-based mission planning, significantly improving computation time.

In summary, our main contributions in this paper are:
1) We propose a reliable HAT scheme using STL and LLM. The LLM translates the human operator's natural language into STL, and the generated STL is then verified through the proposed SFF. The SFF can specify the feasibility of each subformula, i.e., check if the subformula is impossible or unsafe, thereby allowing reliable cooperation between the human and autonomous agents; 2) The proposed SFF can further provide informative feedback to the human operator by utilizing the information on infeasible subformulas. Thus, the proposed framework can explain why the given command was infeasible with detailed reasoning; and 3) We rigorously test the performance of the proposed framework through a set of illustrative numerical simulations and real experiments. 

The rest of the paper is organized as follows. In Section \ref{prelm}, the preliminaries on STL and the reachability analysis are presented. Section \ref{proposed_framework} provides a detailed description of our proposed framework. In Section \ref{sec:experiment}, the results of the numerical simulations and experiments are presented. Lastly, the conclusion is given in Section \ref{sec:conclusion}.


\section{Preliminaries} \label{prelm}

We consider an autonomous system with the following dynamics throughout the paper: 
\begin{equation} \label{dynamics}
    \dot{x}=f(x,t,u,d)
\end{equation}
where $x\in\mathbb{R}^n$ is the state, $u\in\mathbb{R}^m$ is the control input, $d\in\mathbb{R}^l$ is the disturbance, $t$ is the time, and $f(\cdot)$ is the dynamics model. 

\subsection{Signal temporal logic (STL) and decomposition}

STL has the syntax composed of $\text{T}\ |\ \neg\phi \ |\ \phi_1\wedge\phi_2\ |\ \phi_1 \mathcal{U}_{[a,b)}\phi_2$. Here, $\phi_{(\cdot)}$ is STL formulae and $\text{T}$ is \textit{True}. 
$\neg$ is negation, $\wedge$ is conjunction (\textit{and}), and $\mathcal{U}_{[a,b)}\ (a\leq b\in\mathbb{R})$ is the $\textit{until}$ operator. 
Using the above syntax, one can express variety of operators including $\textit{disjunction}\ (\vee, \neg(\neg\phi_1 \wedge \neg\phi_2))$, $\textit{eventually}\ (\Diamond, \text{T}\mathcal{U}_{[a,b)}\phi)$, and $\textit{always}\ (\square, \neg \Diamond_{[a,b)}\neg\phi)$~\cite{leahy2023rewrite}. 

In \cite{leahy2023rewrite}, the methods and conditions to decompose a given STL into simpler shape was investigated:
\begin{align}  
     & \square_{[a,b)}(\phi_1 \wedge \phi_2) \rightarrow_{\square}  \square_{[a,b)}\phi_1 \wedge \square_{[a,b)}\phi_2 \label{split_always} \\ 
     & \phi_1 \mathcal{U}_{[a,b)}\phi_2 \rightarrow_{\mathcal{U}}  \square_{[0,b)}\phi_1 \wedge \Diamond_{[a,b)}\phi_2  \label{split_until} \\ 
     & \Diamond_{[a,b)}(\phi_1 \wedge \phi_2) \rightarrow_{\Diamond}  \square_{[a,b)}\phi_1 \wedge \Diamond_{[a,b)}\phi_2  \label{split_eventually}
\end{align} 
In the above equations, \eqref{split_always} is an exact decomposition, i.e., the decomposed STL is identical with the original STL. Meanwhile, \eqref{split_until} and \eqref{split_eventually} are conservative decompositions, where the original STL is satisfied if the decomposed STL is satisfied, but the opposite is not necessarily true~\cite{leahy2023rewrite}. 

$\textit{Robustness},\ \rho(x,t,\phi)$, is a quantitative semantics that can represent the satisfaction of a given STL \cite{donze2010robust}. $\rho$ is defined as:
\begin{align}
&\rho(x,t,\pi(x(t))\geq c) := \pi(x(t))-c \\
&\rho(x,t,\neg \phi) := -\rho(x,t,\phi) \label{stl_neg} \\
&\rho(x,t,\phi_1 \wedge \phi_2) := min(\rho(x,t,\phi_1),\rho(x,t,\phi_2)) \label{stl_and} \\
&\rho(x,t,\phi_1 \mathcal{U}_{[a,b)}\phi_2) := \nonumber \\ & \max_{t'\in[t+a,t+b]} \min (\rho(x,t',\phi_2),\min_{t''\in[t,t']} \rho(x,t'',\phi_1)) 
\end{align}
The STL $\phi$ is satisfied at state $x$ and time $t$ if $\rho(x,t,\phi)$ is positive. Unless stated otherwise, we write $\rho(x,0,\phi)$ as $\rho_{\phi}(x)$ for the brevity throughout the paper.

\subsection{Backward reachable tube (BRT)}

In this paper, we utilize backward reachable tube (BRT) to check the feasibility of given STL specifications. 
The (maximal) BRT,  $\mathcal{R}^M(t,T;\mathcal{T}(\cdot),\mathcal{C}(\cdot))$, of \eqref{dynamics} is defined as~\cite{chen2018signal}
{\small
\begin{multline} \label{brt}
\mathcal{R}^M(t,T;\mathcal{T}(\cdot),\mathcal{C}(\cdot))=\{x:\exists u(\cdot) \in \mathbb{U}, \forall d(\cdot) \in \mathbb{D}, \exists s \in[t,T], \\ \xi_{x,t}^{u(\cdot),d(\cdot)}(s)\in\mathcal{T}(s), \forall \tau\in[t.s], \xi_{x,t}^{u(\cdot),d(\cdot)}(\tau)\in\mathcal{C}(\tau) \}
\end{multline}}
where $ \xi_{x,t}^{u(\cdot),d(\cdot)}$ is the trajectory starting at $x, t$ and driven by the control $u(\cdot)$ under disturbance $d(\cdot)$, 
$\mathcal{T}$ is the target set, and $\mathcal{C}$ is the constraint set. 

The maximal BRT $\mathcal{R}^M$ is the sublevel set of Bellman value function. In other words, $\mathcal{R}^M(t,T;\mathcal{T}(\cdot),\mathcal{C}(\cdot)) = \{x:h_{\mathcal{R}^M}(t,x) <0\}$. Here, $h_{\mathcal{R}^M}(t,x)$ is defined as $h_{\mathcal{R}^M} (t,x)= \inf_{u(\cdot)\in\mathbb{U}} \sup_{d(\cdot)\in\mathbb{D}} \min_{s\in[t,T]} \max \{h_\mathcal{T} (t, \xi_{x,t}^{u(\cdot),d(\cdot)}(s)),\\  \max_{\tau\in[t,s]} h_\mathcal{C}(\xi_{x,t}^{u(\cdot),d(\cdot)}(\tau)) \}$ for given $h_\mathcal{T}(t,x)$ and $ h_\mathcal{C}(x)$ of target and constraint, respectively~\cite{chen2018signal}.
Similarly, the minimal BRT can be defined as 
{\small
\begin{multline} \label{minimal_brt}
\mathcal{R}^m(t,T;\mathcal{T}(\cdot))=\{x:\forall u(\cdot) \in \mathbb{U}, \exists d(\cdot) \in \mathbb{D}, \exists s \in[t,T], \\ \xi_{x,t}^{u(\cdot),d(\cdot)}(s)\in\mathcal{T}(s)\}.
\end{multline}}

Please refer to \cite{chen2018signal} for more detailed information on computing BRT. Similar to $\rho_{\phi}$, we write $h_{(\cdot)}(0,x)$ as $h_{(\cdot)}(x)$ for brevity throughout the paper.

\subsection{Correspondence between STL and BRT}

Let $\mathcal{S}_{\phi}$ be the states that satisfy $\phi$, i.e., $(\xi_{x,t}^{u(\cdot),d(\cdot)}(\cdot),s) \models \phi \Leftrightarrow \xi_{x,t}^{u(\cdot),d(\cdot)}(s) \in \mathcal{S}_{\phi}$. Then, it is clear that $\mathcal{S}_{\phi}:=\{x :\rho_{\phi}(x)>0 \}$, yielding the following correspondence between the robustness ($\rho$) and Bellman value function of $\mathcal{S}_{\phi}$ ($h_{\mathcal{S}_{\phi}}$)~\cite{chen2018signal}:
\begin{equation} \label{corresp}
    \rho_{\phi}(x) > 0 \Leftrightarrow h_{\mathcal{S}_{\phi}}(x)<0
\end{equation}
Thus, one can compute the set of initial states in which a target system can accomplish a mission described by STL ($\phi$) using reachability analysis \eqref{brt}. 

Moreover, based on \eqref{stl_neg}-\eqref{stl_and} and $\textit{disjunction}\ (\vee, \neg(\neg\phi_1 \wedge \neg\phi_2))$, the followings hold~\cite{chen2018signal}:
\begin{align}
    & \rho_{\phi_1\wedge\phi_2}(x)=min(\rho_{\phi_1}(x), \rho_{\phi_2}(x)) \\ &  h_{\mathcal{S}_{\phi_1\wedge\phi_2}}(x)=max(h_{\mathcal{S}_{\phi_1}}(x),h_{\mathcal{S}_{\phi_2}}(x)) \\
    &  \rho_{\phi_1\vee\phi_2}(x)=max(\rho_{\phi_1}(x), \rho_{\phi_2}(x))\\ &h_{\mathcal{S}_{\phi_1\vee\phi_2}}(x)=min(h_{\mathcal{S}_{\phi_1}}(x),h_{\mathcal{S}_{\phi_2}}(x))
\end{align}
along with the $negation$ ($\neg$): $\rho_{\neg\phi}(x)=-\rho_{\phi}(x)$ and $h_{\mathcal{S}_{\neg\phi}}(x) = -h_{\mathcal{S}_{\phi}}(x)$. 
For $\phi_1 \mathcal{U}_{[t_1,t_2]}\phi_2$, if we define a target set and constraint set as
\begin{align}
    & \mathcal{T}_{\phi_2}=
\begin{cases}
    \{x:(x(\cdot),s) \models \phi_2 \}. & \text{if } t\in[s+t_1, s+t_2] \\
    \emptyset & \text{otherwise, }
\end{cases}\\
& C_{\phi_1} = \{x:(x(\cdot),s) \models \phi_1 \}.
\end{align}
the BRT is defined as $\mathcal{R}^M(t,T,\mathcal{T}_{\phi_2}(\cdot),  C_{\phi_1}(\cdot))$, where one can compute $h_{\mathcal{R}^M}=h_{\phi_1 \mathcal{U}_{[t_1,t_2]}\phi_2}$. Furthermore, from \eqref{corresp}, the robustness $\rho_{\phi_1 \mathcal{U}_{[t_1,t_2]}\phi_2}$ also can be obtained as $\rho_{\phi_1 \mathcal{U}_{[t_1,t_2]}\phi_2} = -h_{\phi_1 \mathcal{U}_{[t_1,t_2]}\phi_2}$.
Please refer to~\cite{chen2018signal} for detail derivation on expressing $until$, $\textit{eventually}\ (\Diamond)$, and $\textit{always}\ (\square)$ operator using BRT.

\subsection{Mixed integer linear programming (MILP) planner }

One can transform a given STL specification to the constraints of MILP~\cite{belta2019formal, raman2014model}. 
Given the linear (or linearized) dynamics of a system, state at time $t$, $x(t)$, and trajectory length $N$; we denote $x(k|t)$ as the $k\in \{1,\dots, N\}$ step ahead prediction of the state and $u(k|t)$ as that of the control input with $x(0|t) = x(t)$.
Here, we define $\mathbf{u}^{N,t} = [u(0,t)^\top, u(1|t)^\top, \dots, u(N|t)^\top]^\top$, and $\mathbf{x}(x(t), \mathbf{u}^{N,t}) = [x(0,t)^\top, x(1|t)^\top, \dots, x(N|t)^\top]^\top$. Then, one can formulate the following model predictive control (MPC) problem with STL specification as constraints
\begin{align} 
\min_{\mathbf{u}^{N,t}} \quad & J (\mathbf{x}(x(t), \mathbf{u}^{N,t}), \mathbf{u}^{N,t}) \label{mpc1}\\
\text{s.t.} \quad
& x(t+1) = A x(t) + B u(t) \label{linear_dyn} \\
& \rho(x(t), t, \phi) \geq 0, \cdots, \rho(x(t+N), t, \phi) \geq 0  \label{mpc3}
\end{align}
where $J(\cdot)$ is the cost function that incorporates mission planning and control objectives.

One can rewrite \eqref{mpc1}-\eqref{mpc3} by properly replacing the predicates as mixed integer constraints ~\cite{raman2014model}. For each predicate $\pi = (a_\pi^\top x(t) \leq b_\pi)$, let $z(t)^{\pi}\in \{0,1\}$ be a variable that represents true if $z=1$ or false if $z=0$. Dropping the time index and parameters for each of the variables for brevity, the following holds:
\begin{align}
    a_{\pi}^\top x - M(1-z^{\pi} + \rho) &\leq b_\pi \\
    a_{\pi}^\top x + M z^{\pi} + \rho &\geq b_\pi.
\end{align}
The conversion of a compound STL formula to MILP constraints is performed recursively and is omitted for brevity. Refer to \cite{belta2019formal, raman2014model} for more details.


\section{Proposed Framework} \label{proposed_framework}
The proposed framework starts by taking a human operator's command in natural language. Then, LLM translates the command to STL which is then verified using the proposed STL feasibility filter (SFF).
Given an STL generated by LLM, the SFF 1) decomposes the STL based on \eqref{split_always}-\eqref{split_eventually}; 2) checks the feasibility of the STL using BRT; and 3) identifies the infeasible STL subformula and generates feedback if the STL turns out to be infeasible, otherwise plans a mission using the MILP-based mission planner. The overall architecture of the proposed framework can be found in Fig.~\ref{intro_fig}.

\subsection{Translate human natural commands into STL}

Assuming a human operator gives a command in natural language text, the proposed framework first translates the human operator's input into STL using LLM. We assume basic environment information, such as a mission map and an obstacle's location, is shared with all human operators, LLM, and SFF. We fine-tune LLMs with several example scenarios to generate the STL with the following outputs: 1) \textit{STL\_formula} includes the translated STL and 2) \textit{atomic\_predicates} includes the dictionary of variables with corresponding predicates. More specifically, \textit{atomic\_predicates} contains the specific positions or constraints of STL syntax as its elements.

\begin{centering}
\begin{tcolorbox}[enhanced, sharp corners=all, colback=gray!5, colframe=black,
  width=0.42\textwidth, boxrule=0.7pt, title=LLM STL generation format, fonttitle=\bfseries]
\texttt{spec = \\
  \{ \text{"STL\_formula":} ...  \\
 \text{"atomic\_predicates":}... \} }
\end{tcolorbox}
\end{centering}

\subsection{STL feasibility filter (SFF)}

\subsubsection{STL Decomposition} The SFF takes the LLM's translation, introduced in the previous section, as its input and checks the feasibility. The SFF first decomposes the LLM's output using \eqref{split_always}-\eqref{split_eventually}.
The decomposition of the operator always ($\rightarrow_{\square}$) induces an identical STL, whereas the other decompositions ($\rightarrow_{\mathcal{U}}$ and $\rightarrow_{\Diamond}$) generate conservative results. In this paper, we define several levels of decompositions as follows: $Level\ 0= \{\rightarrow_{\square} \}, Level\ 1=\{\rightarrow_{\square},\ \rightarrow_{\mathcal{U}} \},$ and $Level\ 2= \{\rightarrow_{\square}, \ \rightarrow_{\mathcal{U}}, \rightarrow_{\Diamond} \}$.

For instance, in case of level 1 decomposition, $\rightarrow_{\square}$ and $\rightarrow_{\mathcal{U}}$ are iteratively applied until all the STL is transformed following \eqref{split_always}-\eqref{split_until}.
As a result of the decomposition, the given STL can be divided into a combination of several individual subformulas separated by the operator \textit{and} ($\wedge$). 
In other words, let $\phi'_i$ be the $i-$th subformula, where $p\in\mathbb{N}$ is the number of subformulas that comprise the decomposed STL ($\phi'$). Then, the decomposed STL can be expressed as $\phi'=\bigwedge^p_{i=1} \phi'_i$. 
Thus, one can deal with simpler STL subformulas ($\phi'_i$) respectively when computing BRTs as opposed to directly handling the nested original STL specifications.

Furthermore, we heuristically confirm from experiments that such decomposition can improve the computation speed for MILP-based mission planning, which will be investigated in further detail in Section~\ref{sec:experiment}.

\subsubsection{Reachability Analysis} 

Given $\phi'=\bigwedge^p_{i=1} \phi'_i$ from the aforementioned decomposition, the SFF computes BRT for each $\phi'_i$ to check the feasibility of STL for the target autonomous system. More specifically, the SFF checks if the \textit{target autonomous system can satisfy $\phi'_i$ during the mission}, i.e., stay within $\mathcal{S}_{\phi'_i}$. 
In BRT perspective, the subformula $\phi'_i$ is feasible for the system if the initial state of the autonomous system ($x_0$) is included in the BRT, i.e., within the set $\{x:h_{\mathcal{S}_{\phi'_i}}(x)<0 \}$.

Once each $\phi'_i$ is examined, the SFF handles \textit{and} operation of two subformulas, $\phi'_1\wedge\phi'_2$ using $max$ operation of corresponding Bellman value functions ($h_{\mathcal{S}_{\phi'_1}}$ and $h_{\mathcal{S}_{\phi'_2}}$) based on~\cite{chen2018signal}.
In other words, the SFF chekcs if $max(h_{\mathcal{S}_{\phi'_1}}(x_0), h_{\mathcal{S}_{\phi'_2}}(x_0))<0$ or equivalently, $min(\rho(x_0,0,\phi'_1), \rho(x_0,0,\phi'_2))>0$ from the robustness' perspective.
Thus, each subformula of the decomposed STL ($\phi'_i$) can be \textit{parallelly} verified with a final \textit{max} (or \textit{min}) operation to determine the feasibility of $\phi'=\bigwedge^p_{i=1} \phi'_i$\footnote{Although this can verify the feasibility of each subformula $\phi'_i$ independently, there may not exist a controller that satisfies $\phi'$ \cite{chen2018signal}. In other words, the current framework can verify the feasibility of each subformula for the target autonomous system, but it may not be able to consider the complex correlation between them (e.g., $\phi_A$ and $\phi_B$ might yield feasible BRT, respectively, even if $\phi_A \wedge \phi_B$ is infeasible). Such a limitation will be resolved by leveraging the multiple reach-avoid problem \cite{chen2025control,chen2025control_arxiv} in future work.}.

If all subformulas are feasible, the proposed framework starts the MILP-based mission planning. Otherwise, if at least one of $\phi'_i$ is infeasible, the given $\phi'$ cannot be performed by the target autonomous system and the mission is rejected. Since the SFF checks the feasibility of each $\phi'_i$, not dealing with $\phi'$ as a whole, one can easily identify infeasible subformulas (i.e., infeasible $\phi'_i$). This information is transferred to LLM to generate informative feedback for the human operator.

\begin{figure}[t]
        \centering
        \includegraphics[width=0.3\textwidth]{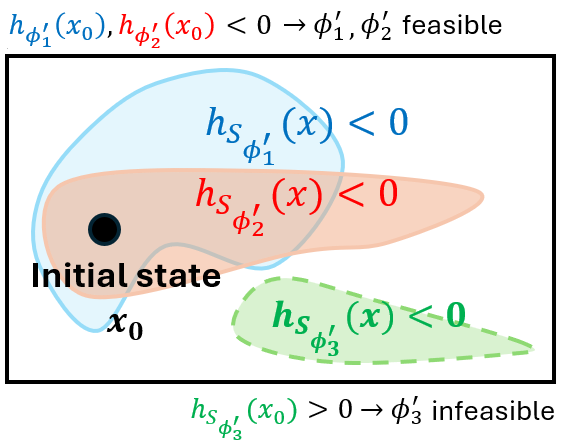}
        \caption{Infeasible subformula identification}
        \label{sff_fig}
\end{figure}

\begin{algorithm}[t]
\caption{Pseudo code of proposed framework}  
\label{alg:all}
\begin{algorithmic}
\State $LLM \gets$ human operator's natural language command
\State $\phi \gets$ generated STL from a LLM
\State \textbf{SFF:}
\State $\phi'=\bigwedge^p_{i=1} \phi'_i \gets$ decomposed STL using \eqref{split_always}-\eqref{split_eventually}
\State Feasibility check using BRT 
\State $\textit{Inf} \gets$ infeasible subformulas
\State \textbf{return} \textit{Inf}
\If{\textit{Inf}$=\emptyset $}
\State \textbf{MILP-based mission planner:}
\State Generate mission using MILP-based mission planner 
\State \textbf{return} Mission trajectory 
\Else
\State \textbf{Feedback generation:}
\State $LLM \gets$\textit{Inf}
\State \textbf{return} Informative feedback
\EndIf
\end{algorithmic}
\end{algorithm}

\subsection{Generating informative feedback}

Once the LLM's translation is verified to be infeasible, the proposed framework rejects the STL and delivers the identified infeasible subformulas to the LLM for informative feedback generation. 
Using such information, along with mission information (e.g., map), the LLM generates feedback to request the human operator to modify the command. An example of LLM's feedback can be found in Fig.~\ref{llm_feedback} in the following section.
Algorithm~\ref{alg:all} shows the pseudo code of the proposed framework.

\section{Experiments}
\label{sec:experiment}

In this section, we validate the effectiveness of our proposed framework using illustrative numerical simulations and real experiments. 
We assume mission information, including the map, is given to all the LLM, SFF, and the human operator, i.e., \textit{the BRTs are computed with obstacle information}.
All of our simulations are performed on a laptop with an i7-13620H processor. The experiments are carried out using the Crazyflie~\cite{giernacki2017crazyflie}, an open-source nano-quadrotor designed for aerial robotics research. The local positions of the Crazyflie quadrotor are measured by the Qualisys motion capture system~\cite{qualisys2024} with a sampling rate of $100$Hz. We implement each of our mission planning scenarios using the \texttt{stlpy} library~\cite{kurtz2022mixed} and the Gurobi solver~\cite{gurobi}, along with \texttt{stlcg} library~\cite{leung2023backpropagation}. The BRT is computed by utilizing \texttt{hj-reachability}~\cite{hj_rs}.

\subsection{Feasibility check for LLM-generated STL} \label{sec51}
In this subsection, the capability of the proposed algorithm to filter out infeasible human commands is demonstrated. To this end, we utilize several LLM models available, ChatGPT-4o~\cite{achiam2023gpt}, Claude 3.7 Sonnet~\cite{anthropic2024claude3}, and Llama 4~\cite{touvron2023llama}. We provide a mission description and map information in natural language, along with the mission map shown in Fig.~\ref{scenario1_2_maps}, and check its feasibility using the proposed SFF. 
Furthermore, we also inquire each LLM to provide its own opinion and reasoning regarding the feasibility of the mission. The prompts and responses of the LLMs can be found in the Appendix.

\subsubsection{Mission Description}

\begin{figure}[h]
    \centering
    \includegraphics[width=0.9\linewidth]{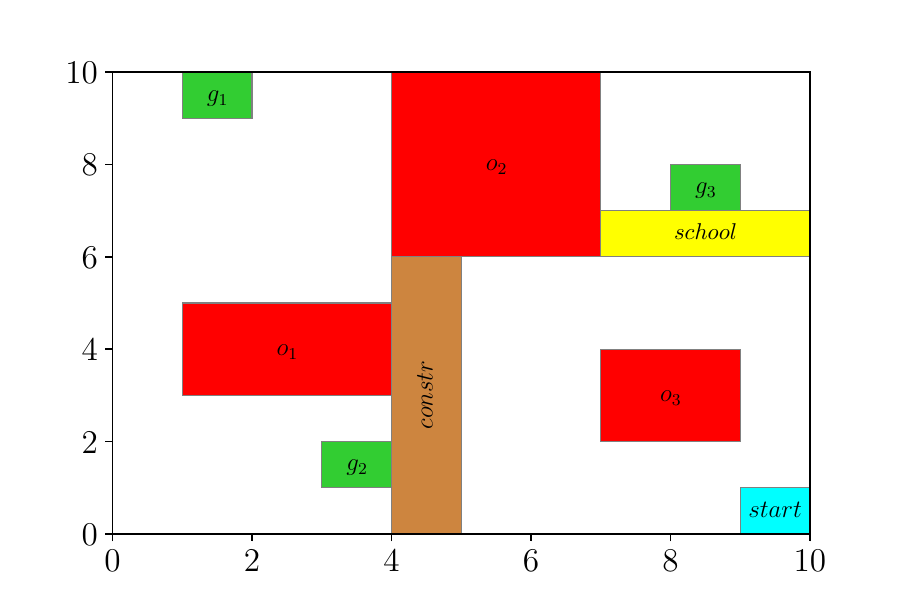}
    \caption{Mission map for Scenario~1}
    \label{scenario1_2_maps}
\end{figure}

Figure~\ref{scenario1_2_maps} shows the mission map of the first scenario. The autonomous agent is set to be a virtual school bus driver, and its mission is to reach the goal points ($g1,\ g2,\ g3$) regardless of its order. We impose time (maximum 60 time steps or equivalently, 30 minutes) constraints for the agent. There are two special areas, the school zone and the construction zone, which have additional contextual constraints. The school zone is closed between 8 to 9 AM; and the construction zone will be closed tomorrow. 
The mission is set to start today at 8:30 AM. Thus, the construction zone does not have any meaning, whereas the school zone is blocked during the mission time, making the mission impossible (the agent cannot reach $g_3$). 

\subsubsection{Simulation Results} 
Table~\ref{Scenario1_result_table} shows the STL generation results from the three LLM models. \cmark \ means the LLM or SFF concludes the given STL is feasible, i.e., the mission can be achieved by the autonomous system; and \xmark \ means infeasible. $\triangle$ means it is non-decisive, where it requests further information for clarification or provides conditional results.
As shown in the table, all LLMs consider the given mission to be (possibly) feasible, although it is not true due to the school zone. On the other hand, the proposed SFF could successfully verify that the given mission is infeasible, catching the flaw in the LLM's translation. Consequently, the proposed framework succeeds in guaranteeing the safety of human-autonomy teaming, preventing infeasible missions from being passed to the MILP-based mission planner. 

\begin{table}[t]
    \centering
    \caption{Self-feasibility check of LLMs and SFF result}
    \label{Scenario1_result_table}
    \resizebox{\columnwidth}{!}{
    \begin{tabular}{lcc|c}
        \toprule
        LLM & Self feasibility & \textbf{SFF} & True \\
        model & check & \textbf{result} & feasibility\\
        \midrule
        ChatGPT-4o        & \cmark      & \xmark   & \xmark  \\
        Claude 3.7 Sonnet & $\triangle$ & \xmark   & \xmark  \\
        Llama 4           & $\triangle$ & \xmark   & \xmark  \\
        \bottomrule
    \end{tabular}%
    }\\[0.25em]
    {\small \quad \quad \quad \quad \cmark\ = feasible, \quad $\triangle$ = nondecisive, \quad \xmark\ = infeasible}
\end{table}

\subsection{Identification of infeasible subformula and informative feedback generation}
\begin{figure}[h]
    \centering
    \includegraphics[width=0.45\textwidth, height=0.5\linewidth]{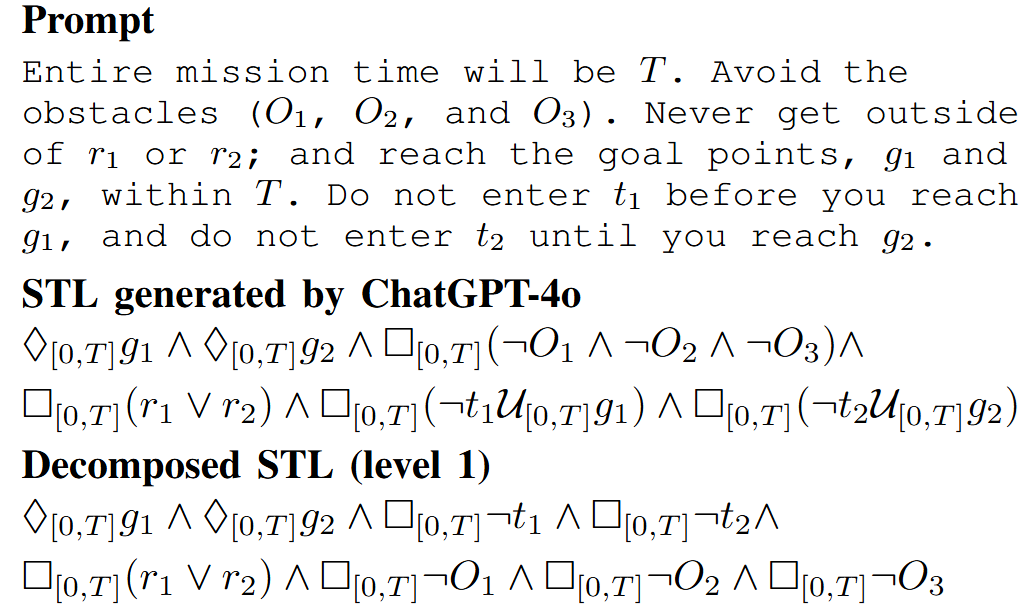} 
    \caption{Mission prompt and corresponding STL generated from LLM}
    \label{scenario1_prompt_stl}
\end{figure}
In this subsection, we demonstrate the general flow of the proposed framework. Additionally, we also demonstrate how the proposed SFF can generate informative feedback to the human operator, based on the identified infeasible subformulas.

\subsubsection{Mission Description}
\begin{figure}[t]
    \centering
    \includegraphics[width=0.5\linewidth]{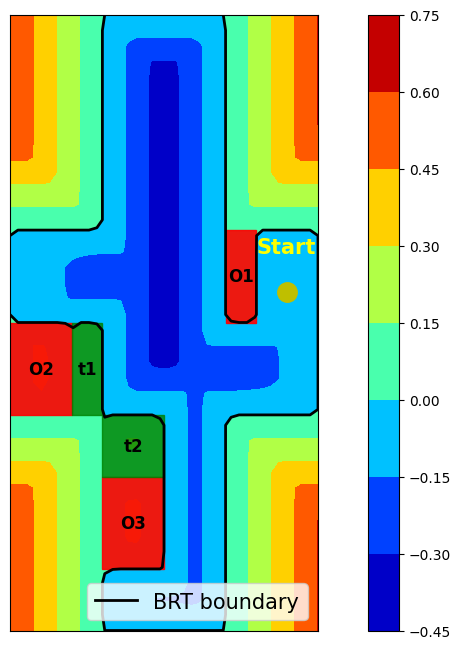}
    \includegraphics[width=0.4\linewidth]{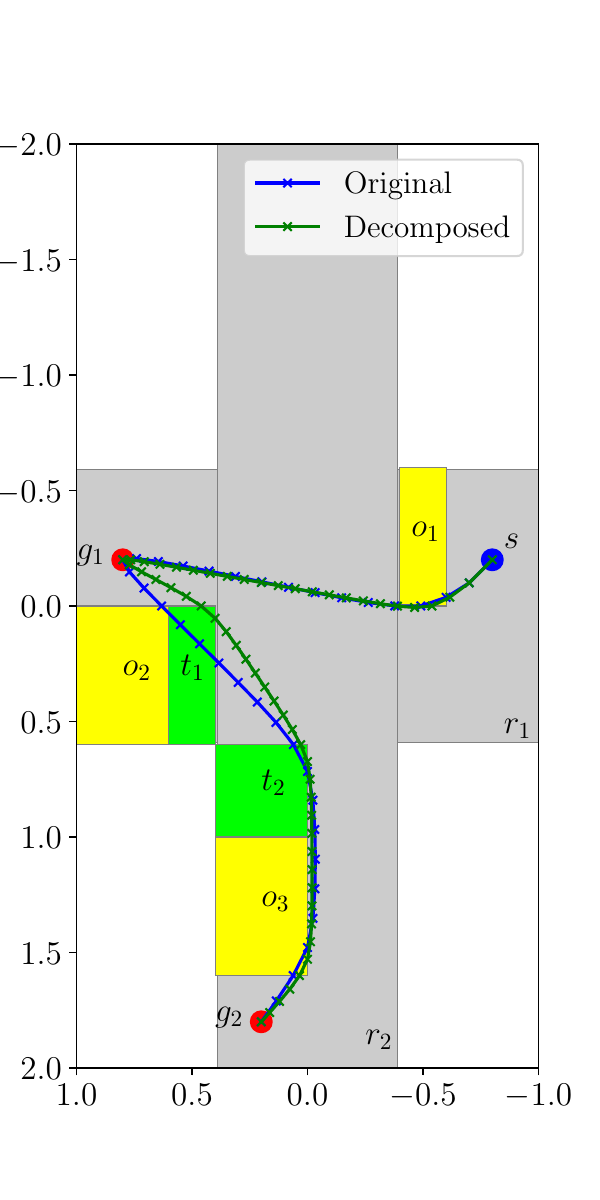}
    \\    
    \includegraphics[width=0.7\linewidth, trim={5.0cm 1.5cm 10.0cm 3.0cm},clip]{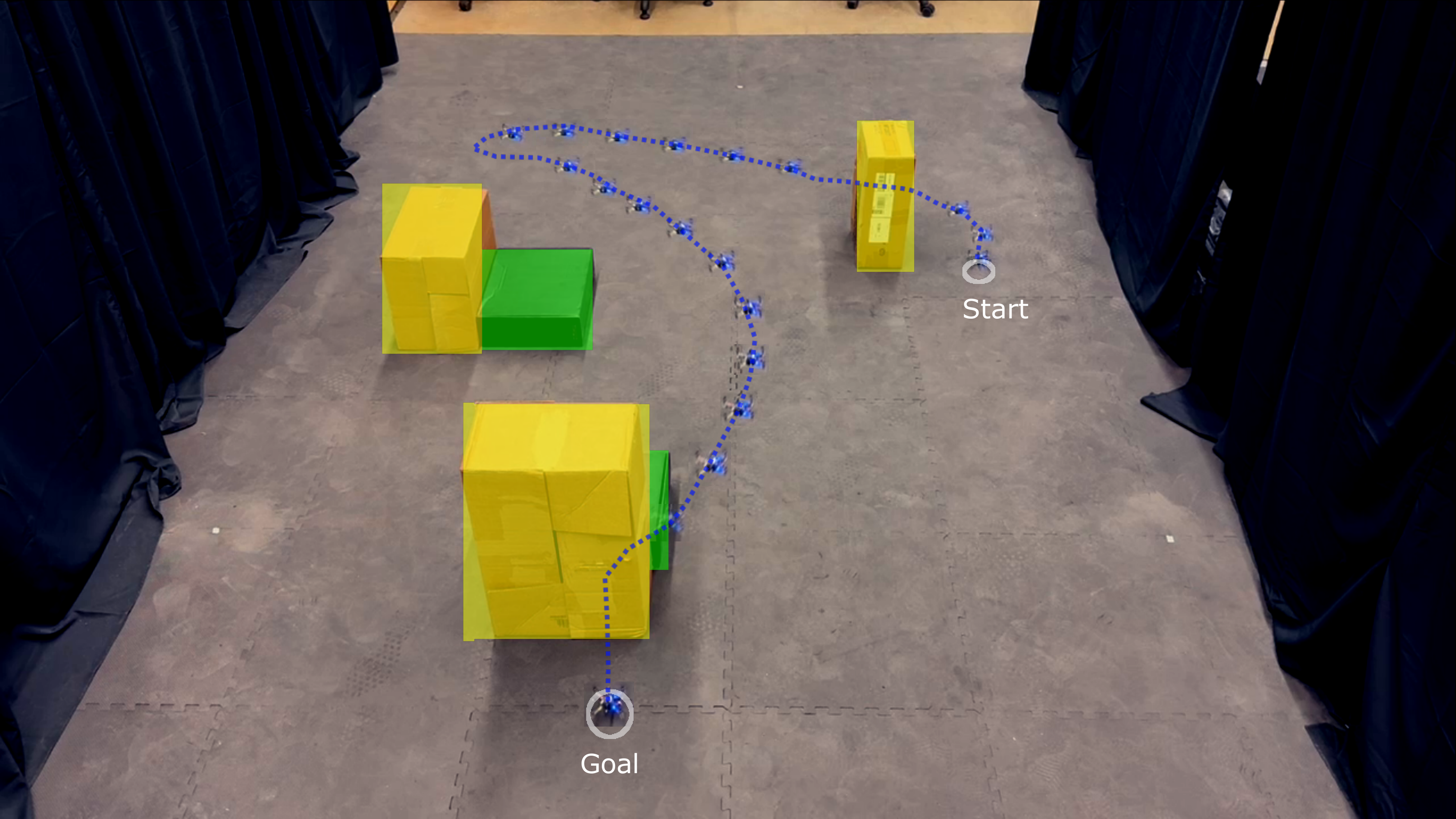}
    \caption{Experiment result of Scenario 2. \textbf{Left:} BRT computed by the SFF. The negative (blue) region represents the initial states from which the autonomous agent can accomplish the mission while satisfying the given STL. \textbf{Right:} The simulated trajectories generated by the MILP-based mission planner. As proven by the SFF, there exists a path that given STL specifications are satisfied. \textbf{Bottom:} Experiment result and trajectory for the decomposed STL (front view).}
    \label{Scenario2_results}
\end{figure}

The natural language prompt and the corresponding generated STLs to perform the mission is given in Fig.~\ref{scenario1_prompt_stl}. As explained in the natural language prompt, the mission is to reach $g_1$ and $g_2$ while staying inside $r_1$ or $r_2$ (gray area). There are obstacles ($O_1-O_3$) and special areas ($t_1$ and $t_2$) where one can enter if the corresponding goal point has been reached. We set the mission time as 30 seconds ($T=30[s]$). 
We generate the STL using ChatGPT-4o and apply level 1 decomposition for SFF as shown in Fig.~\ref{scenario1_prompt_stl}. One can also refer to Fig.~\ref{Scenario2_results} for the mission environment.

\subsubsection{Results From Feasible STL}
Figure~\ref{Scenario2_results} shows the results using the STL presented in Fig.~\ref{scenario1_prompt_stl}. 
In the BRT presented in the figure on the left, the blue (negative) area represents the area where the mission can be accomplished if the initial state is included. As one can easily notice, the autonomous agent starts the mission within the BRT, and thus, the SFF concludes that the given mission is feasible.

The right figure shows the path generated by the MILP-based mission planner, where the blue line represents the path based on the original (LLM's) STL and the red line is that of level 1 decomposition. Although both paths safely complete the mission, they show a slight difference because of the level 1 decomposition's conservativeness. However, as a consequence of the decomposition, we get a significant improvement in terms of the computational time for the mission planning. 
For the original STL specification, it took $ 285.85[s]$ for a $ 30[s]$ mission planning, whereas the decomposed STL yields only $40.67[s]$ for the same mission. 

The figure at the bottom shows the experimental run of the mission using a Crazyflie quadrotor. All calculations were performed on the laptop, and the waypoints are then sent to the Crazyflie over the radio to execute the mission.

\subsubsection{Results From Infeasible STL}
In this experiment, we intentionally modify the decomposed STL shown in Fig.~\ref{scenario1_prompt_stl} to be infeasible; and test if SFF can identify the infeasible subformula.
More specifically, we give the following STL as input to SFF:
\begin{multline} \label{Scenario2_infeasible_STL}
    \Diamond_{[0,T]}g_1 \wedge \textcolor{red}{\Diamond_{[0,T]} (g_2 \wedge r_1)} \wedge \square_{[0,T]} \neg t_1  \wedge  \square_{[0,T]} \neg t_2 \wedge \\
 \square_{[0,T]}(r_1 \vee r_2) \wedge \square_{[0,T]}\neg O_1 \wedge \square_{[0,T]}\neg O_2 \wedge \square_{[0,T]}\neg O_3
\end{multline}
where the modified subformula is marked in red. 
It is clear that in \eqref{Scenario2_infeasible_STL}, the subformula $\Diamond_{[0,T]} (g_2 \wedge r_1)$ is physically impossible, making the mission infeasible. 

\begin{figure}[h]
    \centering
    \includegraphics[width=0.45\textwidth, height=0.24\linewidth]{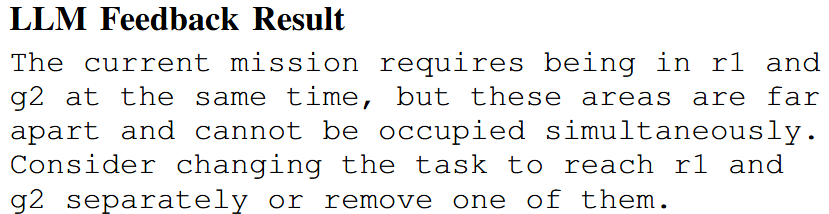} 
    \caption{Feedback generated by LLM leveraging the infeasible STL information}
    \label{llm_feedback}
\end{figure}

From the experiment, the proposed SFF successfully verifies \eqref{Scenario2_infeasible_STL} is infeasible, generating a null BRT (i.e., $\mathcal{R}^M=\emptyset$), and correctly identifies the modified subformula ($\Diamond_{[0,T]} (g_2 \wedge r_1)$) as the only part that induces null BRT. 
Thanks to the identified infeasible subformula from the SFF, LLM can generate more informative feedback that includes the reasoning behind why the mission was infeasible, rather than simply providing boolean information. The generated feedback can be found in Fig.~\ref{llm_feedback}. This result shows that the proposed SFF can facilitate reliable cooperation between a human and an autonomous agent in HAT, realizing informative communication and feedback. 

\subsection{Effect of STL decomposition on mission planning}
In this subsection, we investigate the effect of decomposition on the MILP-based mission planning by comparing the computation time of the original STL and each level of decomposition.

\begin{figure}[t]
    \centering
    \includegraphics[width=0.4\linewidth,
        trim={0.2cm 1.2cm 0.2cm 1.6cm}, clip]{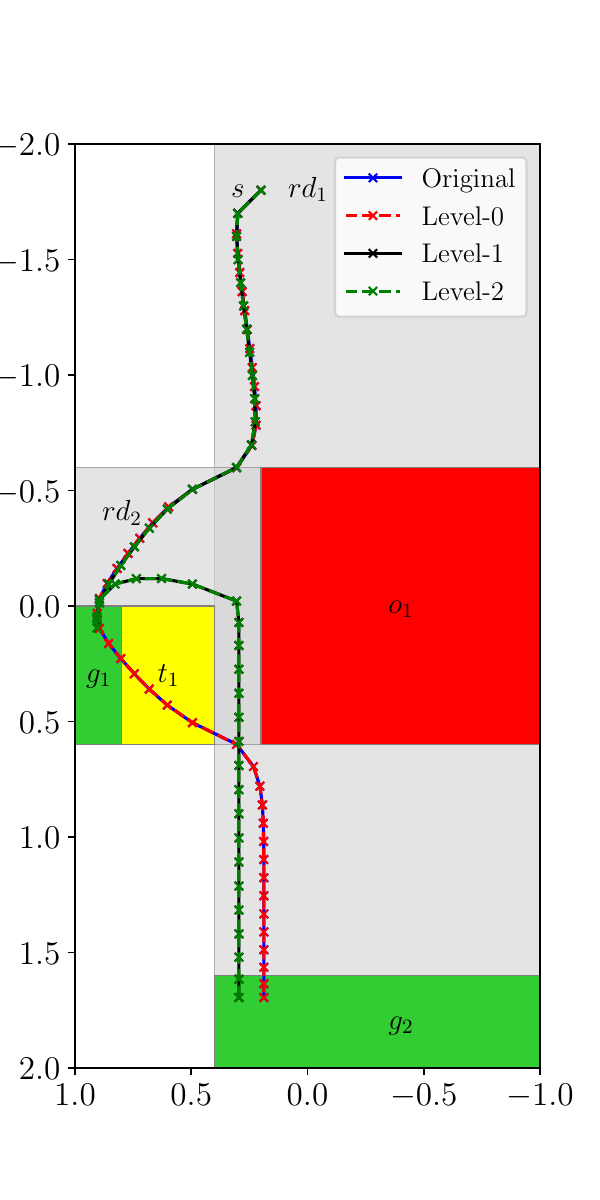}
    \includegraphics[width=0.67\linewidth,
        trim={5.0cm 1.5cm 10.0cm 3.0cm}, clip]{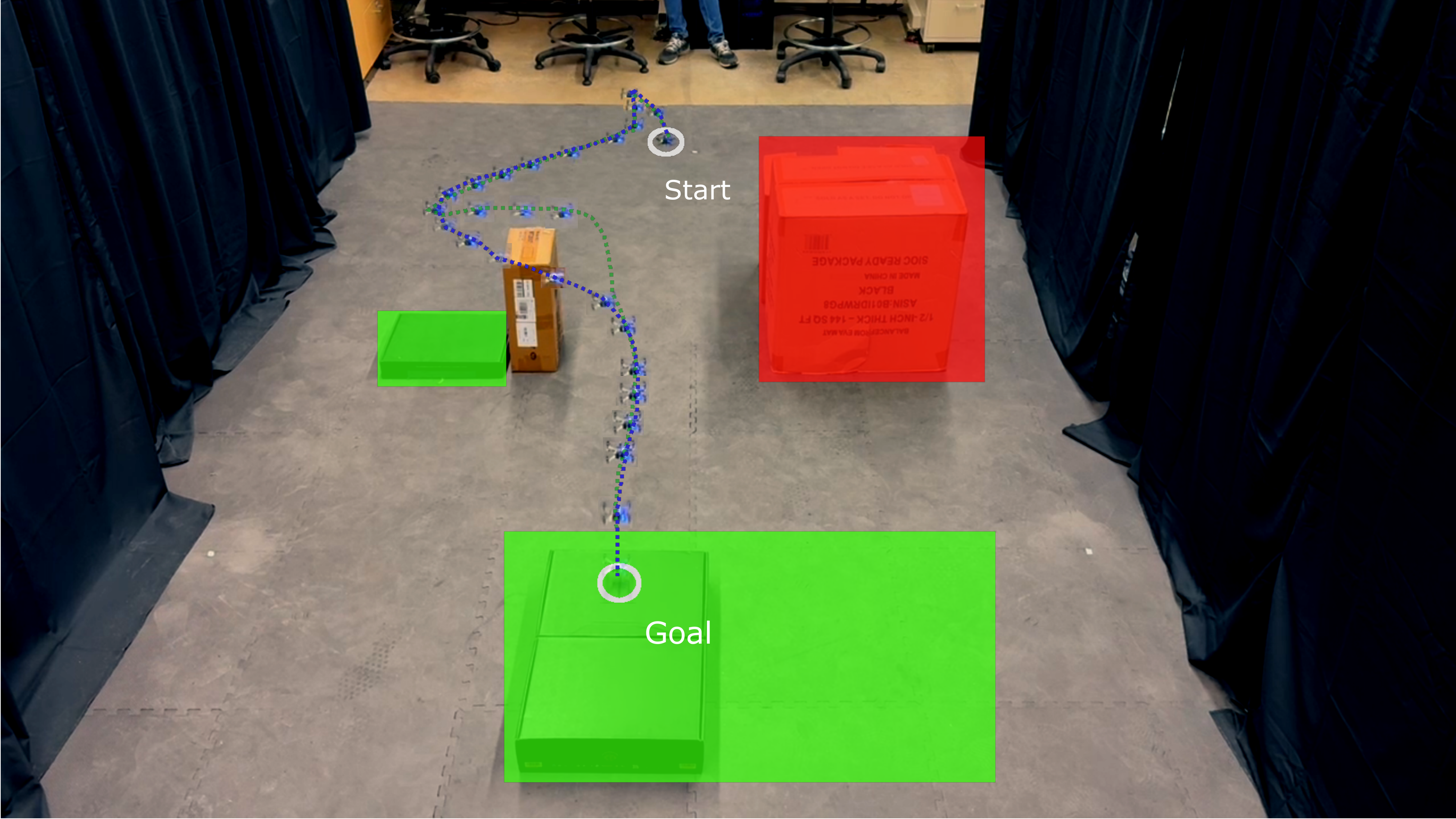}
    \caption{Mission map and experiment results of Scenario~3. \textbf{Top:} simulated trajectories from the MILP-based mission planner using each STL. \textbf{Bottom:} experiment result and trajectories (front view).}
    \label{Scenario3_exp}
\end{figure}

\subsubsection{Mission Description}
The top figure of Fig.~\ref{Scenario3_exp} shows the mission map of Scenario 3. The autonomous agent should reach two goal areas, $g_1$ and $g_2$, while staying within $rd_1$ (vertical rectangle with gray color) or $rd_2$ (horizontal rectangle with gray color). There is an obstacle $O_1$ to avoid, and a special area $t_1$ which can be entered only after reaching $g_1$. Furthermore, the agent must remain at the left half plane ($x>0$) of the map during the entire mission. We manually generate a feasible STL to complete the mission and decompose it according to \eqref{split_always}-\eqref{split_eventually} as follows: 

{\small \begin{align*}
    &\textbf{Original STL: }\square_{[0,T]}(\neg O_1 \wedge \neg Right) \wedge \Diamond_{[0,T]}(Left \wedge g_1) \wedge \\ & \Diamond_{[0,T]}(Left \wedge g_2) \wedge
    \neg t_1 \mathcal{U}_{[0,T]}g_2 \wedge\square_{[0,T]}((Left \wedge rd_1) \vee rd_2)  \\
    &\textbf{Level 0: } \square_{[0,T]}\neg O_1 \wedge \square_{[0,T]} \neg Right \wedge \Diamond_{[0,T]}(Left \wedge g_1) \wedge \\ &  \Diamond_{[0,T]}(Left \wedge g_2) \wedge 
 \neg t_1 \mathcal{U}_{[0,T]}g_2 \wedge\square_{[0,T]}((Left \wedge rd_1) \vee rd_2)  \\
    &\textbf{Level 1: } \square_{[0,T]}\neg O_1 \wedge \square_{[0,T]} \neg Right \wedge \Diamond_{[0,T]}(Left \wedge g_1) \wedge  \\ & \quad  \quad  \quad  \quad  \quad  \quad  \quad  \quad  \Diamond_{[0,T]}(Left \wedge g_2) \wedge \square_{[0,T]}\neg t_1 \wedge \Diamond_{[0,T]} g_2 \wedge  \\ & \quad \quad \quad \quad \quad \quad \quad \quad \quad \quad \quad \quad \quad \quad \square_{[0,T]}((Left \wedge rd_1) \vee rd_2)  \\
    &\textbf{Level 2: } \square_{[0,T]}\neg O_1 \wedge \square_{[0,T]} \neg Right \wedge \square_{[0,T]} Left \wedge  \\ & \Diamond_{[0,T]}g_1 \wedge \Diamond_{[0,T]} g_2 \wedge \square_{[0,T]}\neg t_1 \wedge   \square_{[0,T]}((Left \wedge rd_1) \vee rd_2) 
\end{align*}}
where $Left$ and $Right$ represent the left ($x>0$) and right ($x<0$) half plane of the given map (Fig.~\ref{Scenario3_exp}), respectively.

\subsubsection{Experiment Results}
The top figure of Fig.~\ref{Scenario3_exp} describes the trajectories generated by the MILP-based mission planner using each STL.
One can observe that the level 0 decomposition produces the same trajectory as the original STL due to the exact decomposition scheme. In contrast, the level 1 and 2 decompositions generate slightly different trajectories due to their conservative nature. Nevertheless, all the generated trajectories satisfy the given constraints while performing the mission, and could successfully be implemented in the real experiment as shown in the bottom figure of Fig.~\ref{Scenario3_exp}. 

\begin{table}[t]
    \centering
    \caption{Computation time for mission planning}
    \label{Scenario3_result_table}
    \begin{tabular}{l|cccc}
        \toprule
         & Original & Level 0 & Level 1 & Level 2 \\
        \midrule
        Average    & 54.434 & \textbf{23.923} & 25.163 & 28.632  \\
        Variance   & 1.185  & \textbf{0.212}  & 1.501  & 2.633   \\
        \bottomrule
    \end{tabular}
\end{table}

Meanwhile, Table~\ref {Scenario3_result_table} shows the average computation time of the MILP-based mission planner for the original and each level of decomposed STL. Each value is computed using 100 Monte Carlo simulations. It is clear that the decomposition scheme significantly alleviates the computational burden, improving the computation time by about 50\%. 
Although the level 0 decomposition shows the most efficient planning time among all the cases, levels 1 and 2 also show comparable improvement. 
Considering level 1 and 2 decompositions can facilitate more precise infeasible subformula identification by decomposing the original STL into simpler, smaller pieces; it could significantly benefit reliable and informative communication between humans and autonomous agents.
\section{Conclusion}
\label{sec:conclusion}
This paper introduces an LLM-guided HAT framework that translates a human's natural language command into an STL specification using LLM and validates its feasibility. The proposed SFF filters out infeasible subformulas and generates informative feedback to the human operator. 
The reachability analysis is leveraged to check the given STL, thereby formally verifying the feasibility of a given mission as well as the logical flaws of the translated STL. The decomposition scheme is also shown to improve computational efficiency. The performance of the proposed framework is validated by several numerical simulations and experiments, demonstrating its capability to pre-validate the STL specifications before the planning stage.
As future work, the multiple reach-avoid problem will be accommodated to handle more general STL expressions. Furthermore, a machine learning-based method to compute BRT will also be leveraged to enhance the computational efficiency.



\section*{ACKNOWLEDGMENT}
This material is based on work supported by the National Science Foundation under Grant Number CNS-1836952. Any opinions, findings, conclusions, or recommendations expressed in this work are those of the authors and do not necessarily reflect the views of the National Science Foundation.

\bibliographystyle{IEEEtran}
\bibliography{RAL_bib}

@online{hj_rs,
  howpublished = {\url{https://github.com/StanfordASL/hj\_reachability}}
}

@article{chen2025control_arxiv,
  title={Control Synthesis for Multiple Reach-Avoid Tasks via Hamilton-Jacobi Reachability Analysis},
  author={Chen, Yu and Li, Shaoyuan and Yin, Xiang},
  journal={arXiv preprint arXiv:2509.10896},
  year={2025}
}

@inproceedings{chen2025control,
  title={Control Synthesis for Multiple Reach-Avoid Tasks via Hamilton-Jacobi Reachability Analysis},
  author={Chen, Yu and Li, Shaoyuan and Yin, Xiang},
  booktitle={2025 IEEE 64th Conference on Decision and Control (CDC)},
  pages={5980--5985},
  year={2025},
  organization={IEEE}
}

@article{achiam2023gpt,
  title={Gpt-4 technical report},
  author={Achiam, Josh and Adler, Steven and Agarwal, Sandhini and Ahmad, Lama and Akkaya, Ilge and Aleman, Florencia Leoni and Almeida, Diogo and Altenschmidt, Janko and Altman, Sam and Anadkat, Shyamal and others},
  journal={arXiv preprint arXiv:2303.08774},
  year={2023}
}

@inproceedings{wu2024safety,
  title={On the safety concerns of deploying $\text{llms}$/$\text{vlms}$ in robotics: Highlighting the risks and vulnerabilities},
  author={Wu, Xiyang and Xian, Ruiqi and Guan, Tianrui and Liang, Jing and Chakraborty, Souradip and Liu, Fuxiao and Sadler, Brian M and Manocha, Dinesh and Bedi, Amrit},
  booktitle={First Vision and Language for Autonomous Driving and Robotics Workshop},
  year={2024}
}

@article{cummings2014man,
  title={Man versus machine or man+ machine?},
  author={Cummings, Mary Missy},
  journal={IEEE Intelligent Systems},
  volume={29},
  number={5},
  pages={62--69},
  year={2014},
  publisher={IEEE}
}

@inproceedings{wang2024ensuring,
  title={Ensuring Safety in LLM-Driven Robotics: A Cross-Layer Sequence Supervision Mechanism},
  author={Wang, Ziming and Liu, Qingchen and Qin, Jiahu and Li, Man},
  booktitle={2024 IEEE/RSJ International Conference on Intelligent Robots and Systems (IROS)},
  pages={9620--9627},
  year={2024},
  organization={IEEE}
}

@article{hafez2025safe,
  title={Safe $\text{LLM}$-Controlled Robots with Formal Guarantees via Reachability Analysis},
  author={Hafez, Ahmad and Akhormeh, Alireza Naderi and Hegazy, Amr and Alanwar, Amr},
  journal={arXiv preprint arXiv:2503.03911},
  year={2025}
}

@article{jin2022integrating,
  title={Integrating AI planning with natural language processing: a combination of explicit and tacit knowledge},
  author={Jin, Kebing and Zhuo, Hankz Hankui},
  journal={ACM Transactions on Intelligent Systems and Technology},
  year={2022},
  publisher={ACM New York, NY}
}

@inproceedings{tanneberg2024help,
  title={To help or not to help: $\text{LLM}$-based attentive support for human-robot group interactions},
  author={Tanneberg, Daniel and Ocker, Felix and Hasler, Stephan and Deigmoeller, Joerg and Belardinelli, Anna and Wang, Chao and Wersing, Heiko and Sendhoff, Bernhard and Gienger, Michael},
  booktitle={2024 IEEE/RSJ International Conference on Intelligent Robots and Systems (IROS)},
  pages={9130--9137},
  year={2024},
  organization={IEEE}
}

@article{wang2023conformal,
  title={Conformal temporal logic planning using large language models},
  author={Wang, Jun and Tong, Jiaming and Tan, Kaiyuan and Vorobeychik, Yevgeniy and Kantaros, Yiannis},
  journal={arXiv preprint arXiv:2309.10092},
  year={2023}
}

@inproceedings{gruber2024impact,
  title={The Impact of Transparency on Human-Autonomy Teaming},
  author={Gruber, Mira E and Hancock, PA and Barber, Daniel J and Wohleber, Ryan and Lyons, Joseph B},
  booktitle={Proceedings of the Human Factors and Ergonomics Society Annual Meeting},
  volume={68},
  number={1},
  pages={1783--1788},
  year={2024},
  organization={SAGE Publications Sage CA: Los Angeles, CA}
}

@article{lyons2021human,
  title={Human--autonomy teaming: Definitions, debates, and directions},
  author={Lyons, Joseph B and Sycara, Katia and Lewis, Michael and Capiola, August},
  journal={Frontiers in psychology},
  volume={12},
  pages={589585},
  year={2021},
  publisher={Frontiers Media SA}
}

@article{o2022human,
  title={Human--autonomy teaming: A review and analysis of the empirical literature},
  author={O’neill, Thomas and McNeese, Nathan and Barron, Amy and Schelble, Beau},
  journal={Human factors},
  volume={64},
  number={5},
  pages={904--938},
  year={2022},
  publisher={Sage Publications Sage CA: Los Angeles, CA}
}

@article{matarazzo2025survey,
  title={A Survey on Large Language Models with some Insights on their Capabilities and Limitations},
  author={Matarazzo, Andrea and Torlone, Riccardo},
  journal={arXiv preprint arXiv:2501.04040},
  year={2025}
}

@article{brown2020language,
  title={Language models are few-shot learners},
  author={Brown, Tom and Mann, Benjamin and Ryder, Nick and Subbiah, Melanie and Kaplan, Jared D and Dhariwal, Prafulla and Neelakantan, Arvind and Shyam, Pranav and Sastry, Girish and Askell, Amanda and others},
  journal={Advances in neural information processing systems},
  volume={33},
  pages={1877--1901},
  year={2020}
}

@misc{anthropic2024claude3,
  author       = {Anthropic},
  title        = {Claude 3 Models},
  year         = {2024},
  howpublished = {\url{https://www.anthropic.com/news/claude-3}}
}

@article{wu2024selp,
  title={SELP: Generating safe and efficient task plans for robot agents with large language models},
  author={Wu, Yi and Xiong, Zikang and Hu, Yiran and Iyengar, Shreyash S and Jiang, Nan and Bera, Aniket and Tan, Lin and Jagannathan, Suresh},
  journal={arXiv preprint arXiv:2409.19471},
  year={2024}
}

@article{choi2024data,
  title={Data-Driven Closed-Loop Reachability Analysis for Nonlinear Human-in-the-Loop Systems Using Gaussian Mixture Model},
  author={Choi, Joonwon and Byeon, Sooyung and Hwang, Inseok},
  journal={IEEE Transactions on Control Systems Technology},
  year={2024},
  publisher={IEEE}
}

@article{touvron2023llama,
  title={Llama 2: Open foundation and fine-tuned chat models},
  author={Touvron, Hugo and Martin, Louis and Stone, Kevin and Albert, Peter and Almahairi, Amjad and Babaei, Yasmine and Bashlykov, Nikolay and Batra, Soumya and Bhargava, Prajjwal and Bhosale, Shruti and others},
  journal={arXiv preprint arXiv:2307.09288},
  year={2023}
}

@inproceedings{leahy2023rewrite,
  title={Rewrite-based decomposition of signal temporal logic specifications},
  author={Leahy, Kevin and Mann, Makai and Vasile, Cristian-Ioan},
  booktitle={NASA Formal Methods Symposium},
  pages={224--240},
  year={2023},
  organization={Springer}
}

@article{leung2023backpropagation,
  title={Backpropagation through signal temporal logic specifications: Infusing logical structure into gradient-based methods},
  author={Leung, Karen and Ar{\'e}chiga, Nikos and Pavone, Marco},
  journal={The International Journal of Robotics Research},
  volume={42},
  number={6},
  pages={356--370},
  year={2023},
  publisher={SAGE Publications Sage UK: London, England}
}

@inproceedings{donze2010robust,
  title={Robust satisfaction of temporal logic over real-valued signals},
  author={Donz{\'e}, Alexandre and Maler, Oded},
  booktitle={International Conference on Formal Modeling and Analysis of Timed Systems},
  pages={92--106},
  year={2010},
  organization={Springer}
}

@inproceedings{chen2018signal,
  title={Signal temporal logic meets reachability: Connections and applications},
  author={Chen, Mo and Tam, Qizhan and Livingston, Scott C and Pavone, Marco},
  booktitle={International Workshop on the Algorithmic Foundations of Robotics},
  pages={581--601},
  year={2018}
}

@article{liu2024effect,
  title={Effect of Adaptive Communication Support on Human-AI Collaboration},
  author={Liu, Shipeng and Shrutika, FNU and Zhang, Boshen and Huang, Zhehui and Qian, Feifei},
  journal={arXiv preprint arXiv:2412.06808},
  year={2024}
}

@INPROCEEDINGS{raman2014model,
  author={Raman, Vasumathi and Donzé, Alexandre and Maasoumy, Mehdi and Murray, Richard M. and Sangiovanni-Vincentelli, Alberto and Seshia, Sanjit A.},
  booktitle={53rd IEEE Conference on Decision and Control}, 
  title={Model predictive control with signal temporal logic specifications}, 
  year={2014},
  volume={},
  number={},
  pages={81-87},
  doi={10.1109/CDC.2014.7039363}}

@article{belta2019formal,
  title={Formal methods for control synthesis: An optimization perspective},
  author={Belta, Calin and Sadraddini, Sadra},
  journal={Annual Review of Control, Robotics, and Autonomous Systems},
  volume={2},
  number={1},
  pages={115--140},
  year={2019},
  publisher={Annual Reviews}
}

@misc{qualisys2024,
  author  = {{Qualisys AB}},
  title   = {{Qualisys Motion Capture System}},
  year    = {2024},
  howpublished  = {\url{https://www.qualisys.com}},
}

@inproceedings{giernacki2017crazyflie,
  title={Crazyflie 2.0 quadrotor as a platform for research and education in robotics and control engineering},
  author={Giernacki, Wojciech and Ambroziak, Lukasz and Becker, Miroslaw},
  booktitle={International Conference on Methods and Models in Automation and Robotics},
  pages={37--42},
  year={2017},
  organization={IEEE}
}

@article{kurtz2022mixed,
  title={Mixed-Integer Programming for Signal Temporal Logic with Fewer Binary Variables},
  author={Kurtz, Vince and Lin, Hai},
  journal={arXiv preprint arXiv:2204.06367},
  year={2022}
}

@misc{gurobi,
  author = {{Gurobi Optimization, LLC}},
  title = {{Gurobi Optimizer Reference Manual}},
  year = 2024,
  howpublished = {\url{https://www.gurobi.com}}
}

\clearpage

\section*{APPENDIX}

\subsection{Prompts Used in Scenario 1.} Figure~\ref{scenario1_prompt} shows the prompt used to generate the STL.

\begin{figure}[H]
\centering
\begin{tcolorbox}[enhanced, sharp corners=all, colback=gray!5, colframe=black,
  width=\columnwidth, boxrule=0.5pt, title=Natural language prompt for Scenario 1, fonttitle=\bfseries]
\texttt{Test environment:\\
This image is a test environment where the blue box represents the start node. The red regions denoted by o1([1,4,3,5]), o2([4,7,6,10]) and o3([7,9,2,4]) are the obstacles that have to be avoided at all times. The green boxes are the goal nodes denoted by g1([1,2,9,10]), g2([3,4,1,2]) and g3([8,9,7,8]). There is a construction zone in the region [4,5,0,6] which plans to begin the construction tomorrow. There is also a school zone in the region [7,10,6,7] which is blocked from 8 AM to 9 AM every day. The map is a 10*10 grid with a grid size of 1*1. The X and Y axes also show the coordinates of each grid line. I have mentioned the bounds for each region in a bracket next to them in the form (xmin, xmax, ymin, ymax)\\ \\
Test case:\\
You are an autonomous school driver and need to arrive at all the pick-up (goal) points. You need to come back to the starting point after the mission is completed. The dynamics of the vehicle is double integrator, and the maximum speed is limited to 1 m/s (for both x and y axes). Keep in mind that there are several obstacles there. You depart today at 8:30 AM and need to finish the task within 30 minutes. Assume one time step is 0.5 minutes. \\ \\ 
Now, give me the STL specification to complete this mission. Give the output in the exact specified format. Do not change the format or naming convention at all. Stick to the convention. Check the bounds properly. Please make sure if this mission is feasible, and give me your reasoning why it is feasible (or not)  along with the STL.}
\end{tcolorbox}
\caption{Prompt utilized in Scenario 1 as an input for LLMs}
\label{scenario1_prompt}
\end{figure}

\subsection{Output of LLMs in Scenario 1.} Figures~\ref{scenario1_gpt_stl}-\ref{scenario1_llama} shows the generated STLs from the LLMs and corresponding self feasibility check results. In the figures, $F$, $G$, and $!$ represents $\Diamond$, $\square$, and $\neg$, respectively. $\&$ is $\wedge$ and $|$ means $\vee$.

\begin{figure}[H] 
\centering
\begin{tcolorbox}[enhanced, sharp corners=all, colback=gray!5, colframe=black,
  width=\columnwidth, boxrule=0.5pt, title=STL generated by ChatGPT-4o, fonttitle=\bfseries]
\texttt{
  "$stl\_formula$": "G[0,60](!o1 \& !o2 \& !o3 \& !school) \& G[0,26](!constr) \& F[0,60](g1) \& F[0,60](g2) \& F[0,60](g3) \& F[0,60](start)", \\
  "$atomic\_predicates$": \{ \\
    "o1": "x >= 1 \& x <= 3 \& y >= 4 \& y <= 5", \\
    "o2": "x >= 4 \& x <= 6 \& y >= 6 \& y <= 9", \\
    "o3": "x >= 7 \& x <= 8 \& y >= 2 \& y <= 3", \\
    "school": "x >= 7 \& x <= 9 \& y == 6", \\
    "constr": "x == 4 \& y >= 0 \& y <= 5", \\
    "g1": "x == 1 \& y == 9", \\
    "g2": "x == 3 \& y == 1", \\
    "g3": "x == 8 \& y == 7", \\
    "start": "x == 9 \& y == 0" \}  }
\end{tcolorbox}
\caption{STL generated by ChatGPT-4o in Scenario 1}
\label{scenario1_gpt_stl}
\end{figure}

\begin{figure}[H]
\centering
\begin{tcolorbox}[enhanced, sharp corners=all, colback=gray!5, colframe=black,
  width=\columnwidth, boxrule=0.5pt, title=ChatGPT-4o self feasibility check, fonttitle=\bfseries]
\texttt{feasibility: Feasible,\\
  reasoning: The mission starts at 8:30 AM and must be completed by 9:00 AM (i.e., within 30 minutes or 60 time steps). The school zone is only blocked from 8:00 to 9:00 AM. However, it occupies y=6, x in [7,9], and the goal g3 lies at (8,7), so it is reachable without crossing into the restricted school region. The construction zone starts tomorrow, so it's safe to traverse today. All three goals (g1, g2, g3) and return to start are reachable from the start point at (9,0), and each grid move (1 m) takes 0.5 minutes. So a single-axis diagonal move would take 1 minute (0.5 min per axis), and all targets can be visited within 60 steps. Hence, the mission is feasible.}
\end{tcolorbox}
\caption{Self feasibility check result of ChatGPT-4o}
\label{scenario1_gpt}
\end{figure}


\begin{figure}[t]
\centering
\begin{tcolorbox}[enhanced, sharp corners=all, colback=gray!5, colframe=black,
  width=\columnwidth, boxrule=0.5pt, title=STL generated by Claude 3.7 Sonnet, fonttitle=\bfseries]
\texttt{
  "$stl\_formula$": "G[0,60](!o1 \& !o2 \& !o3 \& !school) \& G[0,60]($vx^2 + vy^2$ <= 1) \& F[0,60](g1) \& F[0,60](g2) \& F[0,60](g3) \& F[0,60](start)", \\
  "$atomic\_predicates$": \{ \\
    "o1": "x >= 1 \& x <= 4 \& y >= 3 \& y <= 5", \\
    "o2": "x >= 4 \& x <= 7 \& y >= 6 \& y <= 10", \\
    "o3": "x >= 7 \& x <= 9 \& y >= 2 \& y <= 4", \\
    "school": "x >= 7 \& x <= 10 \& y >= 6 \& y <= 7", \\
    "g1": "x >= 1 \& x <= 2 \& y >= 9 \& y <= 10", \\
    "g2": "x >= 3 \& x <= 4 \& y >= 1 \& y <= 2", \\
    "g3": "x >= 8 \& x <= 9 \& y >= 7 \& y <= 8", \\
    "start": "x >= 9 \& x <= 10 \& y >= 0 \& y <= 1", \\
    "vx": "dx/dt", \\
    "vy": "dy/dt" \} }
\end{tcolorbox}
\caption{STL generated by Claude 3.7 Sonnet in Scenario 1}
\label{scenario1_claude_stl}
\end{figure}

\begin{figure}[t]
\centering
\begin{tcolorbox}[enhanced, sharp corners=all, colback=gray!5, colframe=black,
  width=\columnwidth, boxrule=0.5pt, title=Claude 3.7 Sonnet self feasibility check, fonttitle=\bfseries]
\texttt{Regarding feasibility: This mission appears to be challenging but potentially feasible. \\
Time constraint: We have 30 minutes (60 timesteps of 0.5 minutes each) \\
Distance analysis: The grid is 10×10 and we need to visit all 3 goals and return to start At maximum speed of 1 m/s, traversing the full diagonal would take about 14.14 seconds Even with an inefficient path visiting all goals, the total distance should be navigable\\ 
Key challenges: The school zone is blocked at our start time (8:30AM), requiring us to avoid [7,10,6,7]. The vehicle must navigate around three obstacles. Double integrator dynamics means acceleration/deceleration must be accounted for.\\
For optimal execution, the route should be carefully planned to visit all goals efficiently while respecting the speed limit and avoiding obstacles. The STL specification captures these requirements within the 60 timestep time horizon.}
\end{tcolorbox}
\caption{Self feasibility check result of Claude 3.7 Sonnet}
\label{scenario1_claude}
\end{figure}


\begin{figure}[H]
\centering
\begin{tcolorbox}[enhanced, sharp corners=all, colback=gray!5, colframe=black,
  width=\columnwidth, boxrule=0.5pt, title=STL generated by Llama4, fonttitle=\bfseries]
\texttt{
  "$stl\_formula$": "G[0,60] (!(o1 | o2 | o3) \& !(school)) \&  F[0,60] (g1 \& F[0,60] (g2 \& F[0,60] g3)) \& F[0,60] (x = 9 \& y = 1)",\\
  "$atomic\_predicates$": \{ \\
    "o1": "(1 <= x <= 4) \& (3 <= y <= 5)", \\
    "o2": "(4 <= x <= 7) \& (6 <= y <= 10)", \\
    "o3": "(7 <= x <= 9) \& (2 <= y <= 4)", \\
    "g1": "(1 <= x <= 2) \& (9 <= y <= 10)", \\
    "g2": "(3 <= x <= 4) \& (1 <= y <= 2)", \\
    "g3": "(8 <= x <= 9) \& (7 <= y <= 8)", \\
    "school": "(7 <= x <= 10) \& (6 <= y <= 7)"
  \} }
\end{tcolorbox}
\caption{STL generated by Llama4 in Scenario 1}
\label{scenario1_llama_stl}
\end{figure}

\begin{figure}[H]
\centering
\begin{tcolorbox}[enhanced, sharp corners=all, colback=gray!5, colframe=black,
  width=\columnwidth, boxrule=0.5pt, title=Llama4 self feasibility check, fonttitle=\bfseries]
\texttt{The mission's feasibility depends on the vehicle's ability to navigate through the grid, avoid obstacles, visit all goal points, and return to the start within the given time frame while respecting the school zone constraint. Given the vehicle's speed and the time available, it seems challenging but potentially feasible depending on the exact start point and the path taken. Precise path planning would be necessary to confirm feasibility.\\
Given the format and requirements, the final answer focuses on the STL specification and the feasibility assessment based on the provided details.}
\end{tcolorbox}
\caption{Self feasibility check result of Llama4}
\label{scenario1_llama}
\end{figure}

\end{document}